\documentclass[twocolumn]{emulateapj}

\usepackage{epsfig}
\usepackage{epsfig}
\usepackage{natbib}
\usepackage{lscape}
\usepackage{enumerate}
\usepackage{multirow}
\usepackage{array}

\def\plotone#1{\centering \leavevmode
\includegraphics[clip=, width=.85\columnwidth]{#1}}

\def\plottwo#1#2{\centering \leavevmode
\includegraphics[width=.45\columnwidth]{#1} \hfil
\includegraphics[width=.45\columnwidth]{#2}}

\newcommand{\cN}[1]{\mathcal{N}}

\def\gsim{\;\rlap{\lower 2.5pt
 \hbox{$\sim$}}\raise 1.5pt\hbox{$>$}\;}
\def\lsim{\;\rlap{\lower 2.5pt
   \hbox{$\sim$}}\raise 1.5pt\hbox{$<$}\;}

\begin{document}


\title{%
Generalized Milankovitch Cycles and Longterm Climatic Habitability
}

 \author{
David S. Spiegel\altaffilmark{1,2},
Sean N. Raymond\altaffilmark{3,4},
Courtney D. Dressing\altaffilmark{1},
Caleb A. Scharf\altaffilmark{5,6},
Jonathan L. Mitchell\altaffilmark{7,8}}
 
 \affil{$^1$Department of Astrophysical Sciences, Princeton
 University, Peyton Hall, Princeton, NJ 08544}
 \affil{$^2$Kavli Institute for Theoretical Physics, UCSB, Santa
   Barbara, CA 93106-4030}
 \affil{$^3$Universit\'e de Bordeaux, Observatoire Aquitain des
   Sciences de l'Univers, 2 rue de l'Observatoire, BP 89, F-33271
   Floirac Cedex, France}
 \affil{$^4$CNRS, UMR 5804, Laboratoire d'Astrophysique de Bordeaux, 2
   rue de l'Observatoire, BP 89, F-33271 Floirac Cedex, France}
 \affil{$^5$Columbia Astrobiology Center, Columbia Astrophysics
 Lab., Columbia University, 550 West 120th Street, New York, NY
 10027}
 \affil{$^6$Department of Astronomy, Columbia University, 550 West
 120th Street, New York, NY 10027}
 \affil{$^7$Institute for Advanced Study, School of Natural Sciences,
 1 Einstein Drive, Princeton, NJ 08540}
 \affil{$^8$Earth and Space Sciences, Atmospheric and Oceanic
   Sciences, and IGPP, UCLA, 595 Charles Young Drive East, Los
   Angeles, CA 90095}
 \vspace{0.5\baselineskip}
 
 \email{
dsp@astro.princeton.edu
}

\begin{abstract}
Although the Earth's orbit is never far from circular, terrestrial
planets around other stars might experience substantial changes in
eccentricity.  Eccentricity variations could lead to climate changes,
including possible ``phase transitions'' such as the snowball
transition (or its opposite).  There is evidence that Earth has gone
through at least one globally frozen, ``snowball'' state in the last
billion years, which it is thought to have exited after several
million years because global ice-cover shut off the carbonate-silicate
cycle, thereby allowing greenhouse gases to build up to sufficient
concentration to melt the ice.  Due to the positive feedback caused by
the high albedo of snow and ice, susceptibility to falling into
snowball states might be a generic feature of water-rich planets with
the capacity to host life.  This paper has two main thrusts.  First,
we revisit one-dimensional energy balance climate models as tools for
probing possible climates of exoplanets, investigate the dimensional
scaling of such models, and introduce a simple algorithm to treat the
melting of the ice layer on a globally-frozen planet.  We show that if
a terrestrial planet undergoes Milankovitch-like oscillations of
eccentricity that are of great enough magnitude, it could melt out of
a snowball state.  Second, we examine the kinds of variations of
eccentricity that a terrestrial planet might experience due to the
gravitational influence of a giant companion.  We show that a giant
planet on a sufficiently eccentric orbit can excite extreme
eccentricity oscillations in the orbit of a habitable terrestrial
planet.  More generally, these two results demonstrate that the
longterm habitability (and astronomical observables) of a terrestrial
planet can depend on the detailed architecture of the planetary system
in which it resides.
\end{abstract}


\section{Introduction}
\label{sec:intro}
Even very mild astronomical forcings can have striking influences on
the Earth's climate.  Although the orbital eccentricity varies between
$\sim$0 and only $\sim$0.06, and the axial tilt, or obliquity, between
22.1$\degr$ and 24.5$\degr$, these slight periodic changes are
sufficient to help drive the Earth into ice ages at regular intervals.
\citet{milankovitch1941} recognized and articulated this possibility
in his astronomical theory of the ice ages.\footnote{Prior to
  Milankovitch, \citet{adhemar1842} and \citet{croll1875} argued that
  astronomical forcing influences glaciation.}  Specifically,
Milankovitch posited a causal connection between three astronomical
cycles (precession -- 23~kyr period, and variation of both obliquity
and eccentricity -- 41~kyr and 100~kyr periods, respectively) and the
onset of glaciation/deglaciation.  Though much remains to be
discovered about the Milankovitch cycles, they are now generally
acknowledged to have been the dominant factor governing the ice ages
of the last several million years \citep{berger1975, hays_et_al1976,
  berger1976, berger1978, berger_et_al2005}.

The nonzero (but, at just 0.04, nearly zero) eccentricity of Jupiter's
orbit is a significant driver of the Earth's eccentricity Milankovitch
cycle (though the effects of the other planets in our solar system are
important, as well).  If Jupiter's eccentricity were much greater, it
could drive larger amplitude variations of the Earth's eccentricity.
This same mechanism might be operating in other solar systems.  Among
the more than 450 currently known extrasolar planets, there are many
that have masses comparable to Jupiter's and that are on highly
eccentric orbits; $\sim$20\% of the known exoplanets have
eccentricities greater than 0.4, including such extreme values as 0.93
and 0.97 (HD~80606: $e=0.93$, \citealt{Naef_et_al_2001,
  Gillon_2009_4}; HD~20782b: $e=0.97$,
\citealt{otoole_et_al2009b}).\footnote{See http://exoplanet.eu}
Furthermore, tantalizing evidence suggests that lower mass terrestrial
planets might be even more numerous than the giant planets that are
easier to detect \citep{otoole_et_al2009b, mayor_et_al2009}.
Therefore, it seems highly likely that many terrestrial planets in our
galaxy experience exaggerated versions of the Earth's eccentricity
Milankovitch cycle.  Furthermore, although the Earth's obliquity
appears always to be contained within a $2.4\degr$
window\footnote{Even these small variations can cause significant
  climatic changes \citep{drysdale_et_al2009}.}, the obliquity of
planets that lack large moons might vary cyclicly or chaotically,
spanning a much larger range of angles \citep{laskar+robutel1993,
  laskar_et_al1993, nerondesurgy+laskar1997}.  Furthermore, if Jupiter
were significantly closer to the Earth (for example, if it were at
3~AU), the Earth's obliquity would vary chaotically even {\it with}
the Moon present \citep{williams1998a, williams1998}.  In addition,
although they are observationally unconstrained, models suggest that
the spin-rates and initial obliquities of exoplanets might assume a
wide range of values \citep{agnor_et_al1999, chambers2001,
  kokubo+ida2007, miguel+brunini2010}.

These kinds of changes in external forcing could have a dramatic
effect on life that requires liquid water.  Since seminal work by
\citet{dole1964} and \citet{hart1979}, a variety of theoretical
investigations have examined the possible climatic habitability of
terrestrial exoplanets.  \citet{kasting_et_al1993} emphasized that the
habitability of an exoplanet depends on the properties of the host
star.  Several authors have considered how a planet's climate and
climatic habitability depend on the properties of the planet, as well
\citep{williams+kasting1997, franck_et_al2000a, franck_et_al2000b,
  williams+pollard2002, gaidos_et_al2005, vonbloh_et_al2007b,
  selsis_et_al2007, dobrovolskis2007, dobrovolskis2009,
  vonbloh_et_al2008, spiegel_et_al2008, spiegel_et_al2009}.  In
particular, two recent works have focused on the climatic effect of
orbital eccentricity.  \citet{williams+pollard2003} used a general
circulation climate model to address the question of how the Earth's
climate would be affected by a more eccentric orbit.  A companion
paper to this one \citep[hereafter D10]{dressing_et_al2010} uses an
energy balance climate model to explore the combined influences of
eccentricity and obliquity on the climates of terrestrial exoplanets
with generic surface geography.  A more eccentric orbit both
accentuates the ratio of stellar irradiation at periastron to that at
apoastron, and increases the annually averaged irradiation (in
proportion to $(1-e^2)^{-1/2}$, where $e$ is eccentricity).  Thus,
periodic oscillations of eccentricity will cause concomitant
oscillations of both the degree of seasonal extremes and of the total
amount of starlight incident on the planet in each annual cycle.
Since these oscillations depend on gravitational perturbations from
other companion objects, the present paper can be thought of as
examining how a terrestrial planet's climatic habitability depends not
just on its star, not just on its own intrinsic properties, but also
on the properties of the planetary system in which it resides.

There is evidence that, at some point in the last billion years, Earth
went through a ``Snowball Earth'' state in which it was fully (or
almost fully) covered with snow and ice (see the review by
\citealt{hoffman+schrag2002} and references therein;
\citealt{williams1993} and \citealt{williams_et_al1998} suggest an
alternative interpretation of the data).  The high albedo of ice gives
rise to a positive feedback loop in which decreasing surface
temperatures lead to greater ice-cover and therefore to further net
cooling.  As a result, the existence of a low-temperature equilibrium
climate might be a generic feature of water-rich terrestrial planets,
and such planets might have a tendency to enter snowball states
\citep{tajika2008}.  The ice-albedo feedback makes it quite difficult
for a planet to recover from such a state \citep{pierrehumbert2005}.
In temperate conditions, the Earth's carbonate-silicate weathering
cycle acts as a ``chemical thermostat'' that tends to prevent surface
temperatures from straying too far from the freezing point of water
\citep{walker_et_al1981, zeebe+caldeira2008}.  A snowball state would
interrupt this cycle.  The standard explanation of how the Earth might
have exited its snowball state is that this interruption of the
weathering cycle would have allowed CO$_2$ to build up to
concentrations approaching $\sim$1~bar over $10^6$-$10^7$ years, at
which point the greenhouse effect would have been sufficient to melt
the ice-cover and restore temperate conditions
\citep{hoffman+schrag2002, caldeira+kasting1992}.  However, an
exoplanet in a snowball state, that is undergoing a large excitation
of its eccentricity, might be able to melt out of its globally frozen
state in significantly less time, depending on the magnitude of the
eccentricity variations and on other properties of the planet.
Exploring this possibility is a primary focus of this paper.

But is it reasonable to expect Earth-like planets around other stars
to have large eccentricities?  The very circular and well-behaved
orbits in the inner Solar System are thought to be the result of
dissipative processes during the late stages of planetary growth, in
particular dynamical friction resulting from interactions between
growing protoplanets and remnant planetesimals
\citep{obrien_et_al2006, morishima_et_al2008, raymond_et_al2006,
  raymond_et_al2009}.  After the relatively short ($\sim$100~Myr)
phase of planet formation, the Solar System's long-term evolution is
thought to have been dominated by dynamical interactions between the
planets. Currently, the Earth's eccentricity oscillates between almost
zero and about 0.06 with a $\sim$100,000 year periodicity
\citep{laskar1988, quinn_et_al1991}.  The dynamics of the Earth's
orbit is controlled by secular forcing from the other planets and
undergoes chaotic evolution on long timescales \citep{laskar1989,
  laskar1990}.  In the context of the known extra-solar planets and a
more general picture of planet formation and evolution, we expect a
much wider range of outcomes than what is seen in the Solar System
\citep{kita_et_al2010}.

During the first few million years of planetary growth, gaseous
protoplanetary disks play an important role in the dynamical evolution
of accreting planets (see review by \citealt{papaloizou+terquem2006}).
Orbital eccentricities of protoplanets may be increased by turbulent
forcing, interactions between protoplanets or with a slightly
elliptical gas disk \citep{papaloizou_et_al2001, kokubo+ida2002,
  oishi_et_al2007, dangelo_et_al2006} and decreased by a combination
of tidal damping and the effects of dynamical friction and aerodynamic
gas drag \citep{adachi_et_al1976, tanaka+ward2004, kominami+ida2004,
  obrien_et_al2006}.  The final eccentricities of a particular system
of terrestrial planets results from a competition between these
numerous effects.  In addition, given that the lifetime of gaseous
disks is far shorter than the Earth's measured accretion timescale
\citep{haisch_et_al2001, pascucci_et_al2006, touboul_et_al2007,
  allegre_et_al2008}, the orbital configuration of any giant planets
in the system will play an important role in the final sculpting of
the terrestrial planets (e.g., \citealt{levison+agnor2003,
  raymond_et_al2004}).

Gap-opening planets ($M_p \gsim M_{\rm Saturn}$;
\citealt{crida_et_al2006}) may acquire eccentricities on the order of
0.1-0.2 from planet-disk interactions \citep{Goldreich_and_Sari_2003,
  dangelo_et_al2006}.  However, these values are too small to explain
the eccentricity distribution of the known extra-solar giant planets,
which is centered at 0.2-0.3 and contains values higher than 0.9
\citep{butler_et_al2006, Wright_et_al2009}.  Several mechanisms have
been proposed to explain the observed eccentricity distribution (see
\citealt{ford+rasio2008} for a summary).  Currently, the leading
candidate is the planet-planet scattering mechanism, which proposes
that most exoplanet systems (75-100\% of them) became dynamically
unstable after their formation, leading to close encounters between
planets and the eventual destruction of some planets
\citep{rasio+ford1996, weidenschilling+marzari1996,
  juric+tremaine2008, thommes_et_al2008}.  Planet-planet interactions
between Earth-sized objects are much more likely to result in
eccentricity-damping collisions than eccentricity-increasing
scattering events \citep{goldreich_et_al2004}.  However, scattering
among giant planets in the vicinity of rocky protoplanets can destroy
the building blocks of Earth-like planets if the giant planets are
close-by, and the protoplanets that survive tend to have large
eccentricities \citep{veras+armitage2006}.  In addition, terrestrial
planets that form in systems with very massive or very eccentric giant
planets tend to themselves have significant eccentricities
\citep{chambers+cassen2002, levison+agnor2003, raymond_et_al2004}.
Simulations of terrestrial planet formation show significant
variations between systems with similar initial conditions (e.g.,
\citealt{raymond_et_al2004, quintana_et_al2007}).  Given that the
range of reasonable input parameters (giant planet orbits, disk
properties) is far wider than that tested with simulations, it seems
reasonable to expect a large diversity in the orbits of extrasolar
Earth-like planets.

The rest of the paper is organized as follows: In \S\ref{sec:model},
we present our energy balance climate model and a dimensional analysis
of its behavior.  In \S\ref{sec:results}, we discuss some scenarios in
which our model suggests that there might be climatic consequences to
eccentricity oscillations that would be interesting from the
perspective of habitability.  In \S\ref{sec:source}, we use secular
perturbation theory and an N-body integrator to investigate what sort
of planetary system architecture might lead to large amplitude
variations in a terrestrial planet's eccentricity.  Finally, in
\S\ref{sec:conc}, we conclude.

\section{Climate Model}
\label{sec:model}
We use nearly the same 1-dimensional time-dependent energy balance
model (EBM) that we have used in previous studies of exoplanet
climatic habitability (\citealt[hereafter SMS08]{spiegel_et_al2008};
\citealt[hereafter SMS09]{spiegel_et_al2009}; see references therein
for a justification of this model in habitability studies).  The
present work's model has a slight modification, described in
\S\ref{ssec:coldstart}, to handle starting in a snowball state.  Our
model treats the redistribution of heat as a diffusive process, and it
(or close variants) has been used by other authors both in the context
of exoplanet habitability \citep[hereafter WK97]{williams+kasting1997}
and in other contexts, including studies of both Martian climate
\citep{nakamura+tajika2002, nakamura+tajika2003} and the Earth's
climate \citep{north_et_al1981}.\footnote{\citet{suarez+held1979} used
  a more sophisticated analog of this EBM, but in an Earth-centric
  context.}  The response of this model to nonzero eccentricity is
explored extensively in a companion paper (D10).  Here, we provide a
brief synopsis of the model.

In accord with WK97 and our own previous work, we use the following
equation for planet surface temperatures ($T$) as a function of time
($t$) and location ($x$):
\begin{equation}
\label{eq:init}
C \frac{\partial T}{\partial t} - \frac{\partial}{\partial x} \left( D(1-x^2) \frac{\partial T}{\partial x} \right) = S(1 - A) - I \, ,
\end{equation}
where $x\equiv\sin\lambda$ and $\lambda$ is latitude.  $S[x,t]$ is the
diurnally averaged stellar irradiation at the latitude band
represented by $x$, on the date represented by $t$.  The values of
heat capacity $C$ and effective diffusivity $D$, and of the albedo and
infrared cooling functions $A$ and $I$ are as described in SMS08 and
SMS09.  We briefly recapitulate. The heat capacity has the following
values over land ($C_l$), ice ($C_i$), and ocean ($C_o$): $C_l =
5.25\times 10^9 \rm~erg~cm^{-2}~K^{-1}$, $C_i = 9.2 C_l$ when
$263<T<273\rm~K$ and $2.0 C_l$ when $T\le 263\rm~K$, and $C_o = 40
C_l$; these values are summarized in
Table~\ref{ta:land_ice_ocean}. The diffusivity $D = 5.394\times 10^2
\rm~erg~cm^{-2}~s^{-1}~K^{-1}$.  The albedo is $A_2$ of SMS08 and
SMS09: $A[T] = 0.525-0.245\tanh[(T-268{\rm~K})/{5\rm~K}]$.  We adopt
two different functional forms for the infrared cooling.  The first is
$I_2$ of SMS08 and SMS09: $I_2[T] = \sigma T^4 / \left\{1+0.5925
(T/273\rm~K)^3\right\}$, where $\sigma = 5.67\times 10^{-5}
\rm~erm~cm^{-2}~s^{-1}~K^{-4}$ is the Stefan-Boltzmann constant.  The
second is $I_{\rm WK97}$ of SMS09 (and WK97), which depends not only
on $T$ but on $p_{\rm CO_2}$, the CO$_2$ partial pressure.  This
function, presented in the appendix of WK97, is a parameterized fit
(third order in $T$ and fourth order in $\ln[p_{\rm CO_2}]$) to
results from a radiative-convective model \citep{kasting+ackerman1986,
  kasting1988, kasting1991}.  We solve equation~(\ref{eq:init}) on a
grid of 145 points equally spaced in latitude.  We use a time-implicit
numerical scheme and an adaptive time-step, as described in SMS08 and
in \citet{hameury_et_al1998}.  Obliquity set to Earth-like 23.5$\degr$
and to $90\degr$, and initial temperature is typically set to be
100~K, with the melting of ice handled as described in
\S\ref{ssec:coldstart}.

\subsection{Dimensional Analysis of EBM}
\label{ssec:nondimen}
As written in equation~(\ref{eq:init}), the energy balance equation
appears similar to a diffusion equation, but is written in terms of a
nondimensional spatial variable ($x$).  It is
instructive to rewrite equation~(\ref{eq:init}) both in fully
dimensional and in fully nondimensional form.

The (1-D, latitudinal) forced heat equation on the sphere may be
written as
\begin{equation}
\label{eq:fully_dim2}
\frac{\partial T}{\partial t} - \frac{K_{yy}}{\cos[s/R]} \frac{\partial}{\partial s} \left\{ \cos[s/R]  \frac{\partial T}{\partial s} \right\} = \frac{1}{C} \left\{S(1 - A) - I\right\} \, .
\end{equation}
Here, $R$ is the planetary radius, $s \equiv R\lambda$ is the distance
north of the equator, and $K_{yy}$ is a dimensionally proper
meridional eddy diffusion coefficient.  This diffussivity may be
expressed as $K_{yy}\equiv R^2/\tau_{\rm diff}$, where $\tau_{\rm
  diff}$ is the diffusion time.  The factors of $\cos[s/R]$ in
equation~(\ref{eq:fully_dim2}) are metric terms that arise from
writing the diffusion equation in spherical coordinates.

Equation~(\ref{eq:fully_dim2}) may be converted to
equation~(\ref{eq:init}) by replacing $s$ with $x\equiv \sin[s/R]$,
multiplying by $C$, and defining $D \equiv C/\tau_{\rm diff}$.  For
the values of $C$ and $D$ quoted above (WK97, SMS08), the diffusion
timescale may be written as
\begin{eqnarray}
\label{eq:difftime}
\nonumber \tau_{\rm diff} & \approx & (C/C_l) (9.7 \times 10^6 {\rm~s}) \\
                         & \approx & (C/C_l)(4 {\rm~months}) \, .
\end{eqnarray}
We find that over land, thick ice, thin ice, and ocean, $\tau_{\rm
  diff}$ has the following respective values: $\sim$4 months, $\sim$7
months, $\sim$3 years, and $\sim$12 years.  The thermal diffusivities
($K_{yy}$) in Table~\ref{ta:land_ice_ocean} ($\sim 10^9
\rm~cm^2~s^{-1}$ for atmosphere above ocean, 40 times greater for
atmosphere above land) might seem surprisingly large for Earth; in
fact, however, they comport with what is expected when a process that
involves large-scale advective motions is modeled as being purely
diffusive \citep{kao+kau1980,
  keeling+heimann1986}.\footnote{\citet{lorenz1979} justified the
  diffusive approximation for studies of large scale climatic trends;
  this treatment has been used in many EBMs in the geophysical
  literature in the last several decades (e.g.,
  \citealt{suarez+held1979, north_et_al1981, north_et_al1983}).
  Furthermore, \citet{showman_et_al2009b} provide an illuminating
  discussion of the applicability of the diffusive approximation.  We
  explored the consequences of varying $D$ in SMS08, SMS09, and D10.}

Alternatively, we may write a fully nondimensional version of
equation~(\ref{eq:init}) by mapping $t \mapsto P_{\rm orb} t^*$, $T
\mapsto T_0 T^*$, $I \mapsto I_0 I^*$, and $S \mapsto I_0 S^*$, with
$I_0 \equiv \sigma {T_0}^4$ ($\sigma$ is the Stefan-Boltzmann
Constant).  Here, $P_{\rm orb}$ is the orbital period and $t^*$ is a
dimensionless time variable; $T_0$ is a typical temperature and $T^*$
is a dimensionless temperature variable; $I_0$ is the blackbody
cooling rate associated with temperature $T_0$ and $I^*$ is a
dimesionless infrared cooling variable; and $S^*$ is a dimensionless
insolation variable.  Noting that the radiative timescale, or thermal
inertia, may be written $\tau_{\rm rad}\equiv C T_0 / I_0 = (C/\sigma)
{T_0}^{-3}$, we are left with the following nondimensional form of the
equation:
\begin{equation}
\label{eq:nondim}
\frac{\partial T^*}{\partial t^*} - \mathcal{K}^* \frac{\partial}{\partial x} \left\{ \left(1-x^2\right) \frac{\partial T^*}{\partial x} \right\} = \mathcal{R}^* \left\{S^*(1 - A) - I^*\right\} \, .
\end{equation}
Here, $\mathcal{K}^* \equiv P_{\rm orb}/\tau_{\rm diff}$ may be
thought of as a dimensionless thermal diffusivity, while
$\mathcal{R}^* \equiv P_{\rm orb}/\tau_{\rm rad}$ may be thought of as
a nondimensional ``thermal elasticity.''  If $\mathcal{R}^* \gg 1$ the
planet's temperature will tend to be determined by the instantaneous
irradiation; conversely, if $\mathcal{R}^* \ll 1$, the temperature
will change little over the course of an annual cycle.  Writing the
equation as in (\ref{eq:nondim}) makes clear that increasing the
orbital period (i.e., increasing the semimajor axis $a$ or decreasing
the stellar mass at fixed luminosity) increases the model planet's
ability both to redistribute thermal energy in a given fraction of a
year, and to approach radiative equilibrium.  While this is
intuitively obvious, it is important to notice that, even while
holding many of a model planet's dimensional parameters constant (in
particular, while holding $C$ and $D$ fixed), if the orbital
separation changes then the nondimensional diffusivity and elasticity
change as well.  This is another way of looking at an issue that was
explored in SMS08: in short, timescales matter, and a closer planet is
not simply a more strongly irradiated version of a more distant
planet.

\subsection{Modeling a Cold Start}
\label{ssec:coldstart}
We place a frozen planet in a variety of pre-set orbits in order to
explore the capacity for the orbital configuration (specifically, the
semimajor axis, the eccentricity, and the obliquity) to thaw such a
planet.  In previous work using this model, we have set ``hot start''
initial conditions in which the initial temperature is set far above
the freezing point of water.  In modeling the melting of a Snowball
Earth planet, we are in a different regime.  We have previously
implicitly assumed that the latent heat involved in melting ice and
freezing water is negligible.  Is this assumption still valid in the
case of a snowball planet that might have a layer of ice that is a
kilometer or more thick?

The latent heat of melting ice ($L_{\rm ice} = 3.3\times 10^9
\rm~erg~g^{-1}$) may be considered negligible if it is small compared
with the mean specific energy that is deposited into ice in a year,
$\left<dE/dm\right>$.  For a crude upper bound on this quantity, we
consider just incident radiant energy:
\begin{equation}
\left<\frac{dE}{dm}\right> \lsim \frac{P_{\rm orb} F_{\rm rad}}{h \rho_{\rm ice}} \, ,
\label{eq:Edep}
\end{equation}
where $F_{\rm rad}$ is the typical incoming radiative flux on a planet
around a Sun-like star, $h$ is the height of a layer of ice, and
$\rho_{\rm ice}$ is the density of ice.  At a distance $a$ around a
Sun-like star, $P_{\rm orb} = (1{\rm~yr})(a / 1{\rm~AU})^{3/2}$, and
the typical stellar irradiation flux is
\begin{eqnarray}
\nonumber F_{\rm rad} & \sim & q_0/4 \left( \frac{a}{1\rm~AU} \right)^{-2}\\
\label{eq:radflux}
                      & = & 3.4 \times 10^5 \left( \frac{a}{1\rm~AU}
                      \right)^{-2} \rm~erg~cm^{-2}~s^{-1},
\end{eqnarray}
where $q_0$ is the solar constant, the flux at the subsolar point on
Earth.  Therefore, $L_{\rm ice} \not< \left< dE/dm \right>$ when
\begin{eqnarray}
\nonumber h & \gsim & \frac{P_{\rm orb} F_{\rm rad}}{\rho_{\rm ice} L_{\rm ice}} \\
 & \sim & (35 {\rm~m}) \left( \frac{a}{1\rm~AU}\right)^{-1/2} \, .
\end{eqnarray}
Since the thickness of ice on Earth varies seasonally by far less than
35~m \citep{semtner1976}, it is not surprising that SMS08 and SMS09
found excellent agreement between our EBM and the Earth's seasonal
variations of temperature (SMS08) and radiative and diffusive fluxes
(SMS09).  A planet at 1~AU from a Sun-like star, however, does not
receive enough radiant energy in a year to melt a layer of ice that is
more than 35~m thick.

We here adopt a simple algorithm to handle the melting of an ice
layer.\footnote{See \citet{semtner1976} for a more sophisticated model
  of the vertical diffusion of heat through an ice layer.}  At a given
location, if the temperature has never reached 273~K, the albedo is
set to its maximum value (0.77), to indicate that the surface is still
covered with snow and ice.  When the local temperature reaches 273~K,
further excess of radiant/diffusive energy (per area) of $\Delta E_A$
is treated as changing the thickness of the local ice layer by $\Delta
h$, where $\Delta h = -\Delta E_A / (\rho_{\rm ice} L_{\rm ice})$.
During such periods of melting, the albedo reduces to 80\% of its
maximum value ($A=0.616$) because snow and ice are less shiny when
melting than when fully frozen \citep{koltzow2007, mellor+kantha1989,
  curry_et_al2001}.  Consider a patch of surface that has just been
melting but, because of the onset of winter, now receives less
incident flux than its radiant and diffusive losses.  First, the
summer's melt-water re-freezes ($\Delta E_A$ is negative, so $\Delta
h$ is positive) while the temperature remains fixed at 273~K.  Once
all melt-water has re-frozen, subsequent energy deficits result in the
temperature decreasing, as specified by equation~(\ref{eq:init}).

Instead of assuming an initial ice thickness and integrating until all
of the ice has melted, we recognize that if a portion of a model
planet, during a given year, receives more incident radiant energy
than its radiant and diffusive losses, then it will continue to
receive an excess of energy in all following years.  Therefore, we
define a patch of surface as having ``melted'' once there has been
positive ``net melting'' over the course of a year (i.e., the summer's
melt-water does not fully refreeze during the winter).  In this way,
we compress what would be perhaps $\sim$10$^3$ years of melting on a
real planet to a single year.\footnote{The estimate of $\sim$10$^3$
  years comes from estimating the length of time required to melt a
  $\sim$kilometer-thick ice layer at $\sim$100~cm~yr$^{-1}$, though
  either the thickness or the rate of melting could differ
  significantly from these estimates.}  If an initially frozen model
planet's surface has partially melted within the integration time of
130 model years, we note this model's orbital configuration as one
that could melt a planet out of a snowball state.  Our simple
algorithm can be criticized for a variety of reasons (e.g., perhaps
the thickness of the ice ought to be constrained to be continuous or
differentiable) but, given other uncertainties involved in modeling
the climate of exoplanets, ours is a reasonable first ansatz to
explore the principle under consideration in this paper.\footnote{See
  WK97 for an extensive discussion of the utility and limitations of
  energy balance climate models in the context of exoplanet climate
  modeling.}

Finally, over the $\sim$$10^5$ years that an exo-Earth's eccentricity
Milankovitch cycle might last, a globally frozen planet that is
comparably geologically active to the Earth might release
$\sim$0.01~bars of CO$_2$ due to volcanism (WK97;
\citealt{holland1978}).  With this in mind, we examine our model's
behavior with three different infrared cooling functions: (i) $I_2$ of
SMS08 and SMS09, which is for an Earth-like greenhouse gas abundance
($\sim$$3\times 10^{-4}$~bars CO$_2$); (ii) $I_{\rm WK97}$ of WK97,
with the partial pressure of CO$_2$ set to 0.01~bars; and (iii)
$I_{\rm WK97}$ with variable CO$_2$ partial pressure.  Once a
cold-start model's ice-cover has completely melted somewhere, silicate
weathering might recommence, thereby removing CO$_2$ from the
atmosphere.  On a real planet, the rate at which melting continues
would involve a complicated interplay between the thickness of the ice
cover, the rates of both CO$_2$ sequestration (through weathering) and
release (through volcanism), and the rate of change of astronomical
forcing.  Since none of these values are constrained for generic
exoplanets, we instead adopt a crude version of a ``chemical
thermostat'' for cooling function (iii): $p_{\rm CO_2} =
10^{-2-(T-250)/27}$~bars if $250 {\rm~K}<T<290$~K; $p_{\rm CO_2} =
3.3\times 10^{-4}$~bars if $T\ge 290$~K; and $p_{\rm CO_2} =
0.01$~bars if $T\le 250$~K.\footnote{The work of
  \citet{lehir_et_al2009} suggests that it might take significantly
  longer than we assume for weathering processes to reestablish
  equilibrium CO$_2$ concentrations after a snowball event.}  In
models in which the CO$_2$ concentration is adjusted with temperature,
this adjustment begins after the the temperature has somewhere
exceeded 273~K for an entire model year.

\section{Climate Results}
\label{sec:results}
When a terrestrial planet's orbital eccentricity or obliquity varies
with time, a variety of climate effects are possible.  The climate
simulations described in D10, \citet{williams+pollard2002}, and
\citet{williams+pollard2003} give us a rough idea of how the climates
of some planets with time-variable astronomical forcing might behave.
But if a planet undergoes a phase transition -- entering into a
snowball state, or melting out of one, the effect of varying
eccentricity is less clear.

Here, we address two questions.  (i) What eccentricity is required to
melt out of a snowball state?  (ii) How does an oscillating
eccentricity (a Milankovitch-like cycle) affect the climate of a
planet that has slipped into a snowball state?

\subsection{Cold-Start Planets}
\label{ssec:coldstart_results}
Rather than attempting to map fully the region of parameter space that
leads to melting planets, we instead explore the eccentricity required
to melt a cold-start model in several illustrative examples.  In
particular, we consider snowball planets at 0.8~AU and at 1.0~AU.

For planets with nonzero eccentricity, two angles are required to
describe the direction of the spin-axis with respect to the orbital
ellipse.  D10 defines these as two obliquity angles: the ``polar
obliquity'' ($\theta_p$), which is the familiar angle between the spin
axis and the orbital plane's normal, and the ``azimuthal obliquity''
($\theta_a$), which is the difference in true anomaly from the
periastron point to the northern winter
solstice.\footnote{Equivalently, it is the angle between the vector
  from the star to the periastron point and the projection of the
  planet's spin-vector onto the orbital plane.}  D10 suggests that for
planets with high eccentricity, azimuthal obliquity angles near
$30\degr$ might be somewhat more stable against global freeze-over.
Taking our cue from this result, we set the azimuthal obliquity angle
to $30\degr$ for all models considered here.  In our brief survey of
parameter space, we consider two values of polar obliquity: Earth-like
$23.5\degr$, and an extreme value of $90\degr$.  We note that, if an
exoplanet experiences Milankovitch-like variations of precession and
obliquity (and in particular if its polar obliquity is not stabilized
by a moon), then its azimuthal and polar obliquity angles might be
expected to swing through a large portion of all possible values.

\begin{figure}
\epsscale{1.0}
\centerline{\plotone{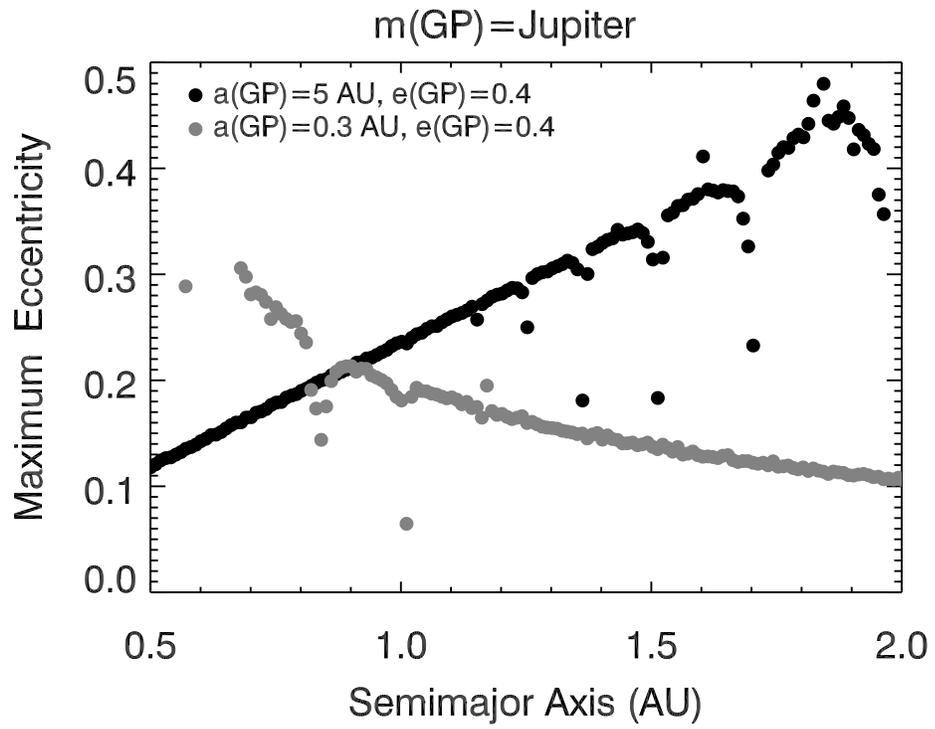}}
\caption{The maximum eccentricity reached by a disk of massless test
  particles over a million year integration of two systems containing
  a single Jupiter-mass giant planet in different configurations:
  $a_{GP} = 5 {\rm AU}, e_{GP} = 0.4$ (black dots), and $a_{GP} = 0.4
  {\rm AU}, e_{GP} = 0.2$ (gray dots).}
\label{fig:emax}
\end{figure}

The eccentricities required to melt out of a global snowball state are
shown for each of 8 model planets in Table~\ref{ta:required_ecc} (two
semimajor axes $\times$ two polar obliquities $\times$ two cooling
functions).  Not surprisingly, the eccentricity need not reach values
that are as great if the greenhouse effect is enhanced by greater
atmopsheric CO$_2$ content.  The values listed (ranging between 0.29
and 0.86) may all be excited by Milankovitch-like cycles in other
solar systems (see Figs.~\ref{fig:e-t} and \ref{fig:emax}).  If the
geochemistry of an exoplanet should cause either its ice to have lower
albedo than we assumed, or its volcanos to release CO$_2$
significantly more rapidly than we considered, the required
eccentricity values might be even lower.  Furthermore, it is likely
that a real planet would have additional feedback processes that might
either amplify or damp the warming caused by increased eccentricity.

\begin{figure}
\centerline{\plottwo{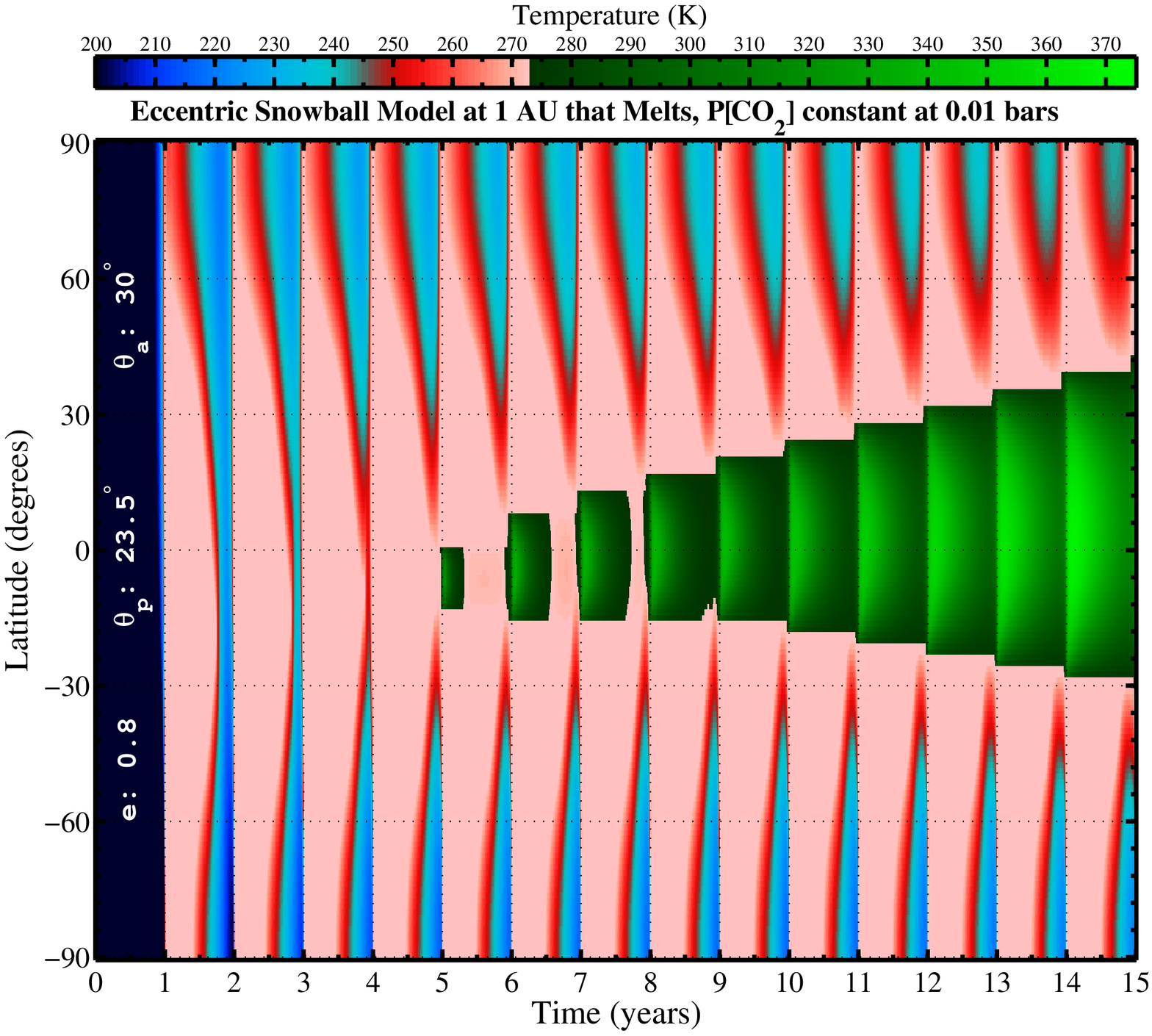}{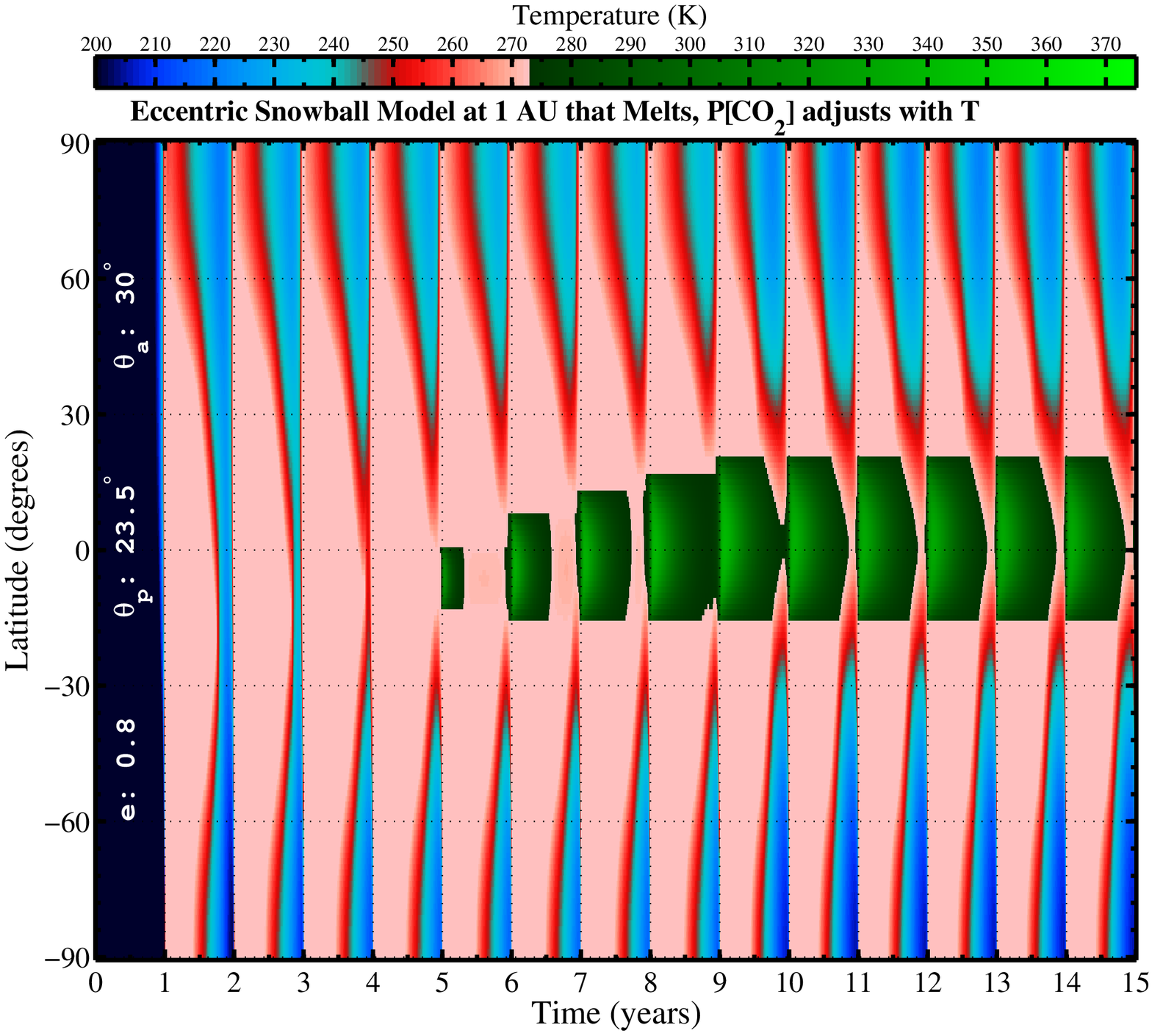}}
\caption{Temperature evolution maps for cold-start models at 1~AU.
  Both models have orbital eccentricity of 0.8 along with Earth-like
  $23.5\degr$ polar obliquity and 1~bar surface pressure. Temperature
  is initialized to 100~K, and quickly rises to near 273~K.  The
  melting of the ice-cover is handled in accordance with the
  prescription of \S\ref{ssec:coldstart}.  {\bf Left:} CO$_2$ partial
  pressure is held constant at 0.01~bars.  In this model, once the
  equatorial region melts, the region of surface that has melted
  ice-cover grows steadily until the entire planet has melted, and
  temperatures eventually grow to more than 400~K over much of the
  planet (not shown).  {\bf Right:} CO$_2$ partial pressure varies
  with temperature, in a crude simulation of a ``chemical
  thermostat''.  In this model, the climate reaches a stable state
  with equatorial melt regions and polar ice-cover.}
\label{fig:tempmap}
\end{figure}

Figure~\ref{fig:tempmap} provides a graphical demonstration of the
temperature evolution on cold-start planets.  Both the left and right
panels depict the climate evolution of models with polar obliquity of
$23.5\degr$, azimuthal obliquity of $30\degr$, and eccentricity of
0.8.  Both models use the WK97 infrared cooling function.  In the left
panel, the CO$_2$ partial pressure is held constant at 0.01~bars.  In
the right panel, the CO$_2$ partial pressure begins at 0.01~bars but
then, a year after there is positive net melting somewhere on the
planet, it adjusts with temperature as described in
\S\ref{ssec:coldstart}.  For the first $\sim$9 years, both panels
evolve identically.  After $\sim$9 years, an equatorial region of
year-round temperate conditions develops.  The model in the left panel
at this point continues to warm progressively.  Once the equatorial
region melts, the fraction of surface that has melted ice-cover grows
steadily until the entire planet has melted, and temperatures
eventually grow to more than 400~K over much of the planet (not
shown).  In the right panel, in contrast, the thermostat begins to
operate after $\sim$9 years, causing the CO$_2$ concentration to
decrease significantly.  The negative feedback causes this model to
achieve a stable equilibrium climate.

We demonstrate in \S\ref{sec:source} that, depending on system
architecture, large variations of eccentricity are possible on
timescales of less than 10$^6$~years.  In this context,
Table~\ref{ta:required_ecc} and Fig.~\ref{fig:tempmap} make it clear
that there are possible circumstances in which eccentricity excitation
could dramatically warm an initially frozen planet.  Intriguingly, in
order for a planet to melt out of a snowball state by virtue of
increased eccentricity and then to remain with a stable, habitable
climate, it might have to have released some greenhouse gas during its
frozen period.  In other words, the classical explanation for how the
Earth might have exited its own snowball state might need to apply to
some extent, although the increased eccentricity can significantly
reduce how much greenhouse gas is required.

\subsection{Experimenting with a Pseudo-Milankovitch Cycle}
\label{ssec:milank}
\begin{figure}
\centerline{\plottwo{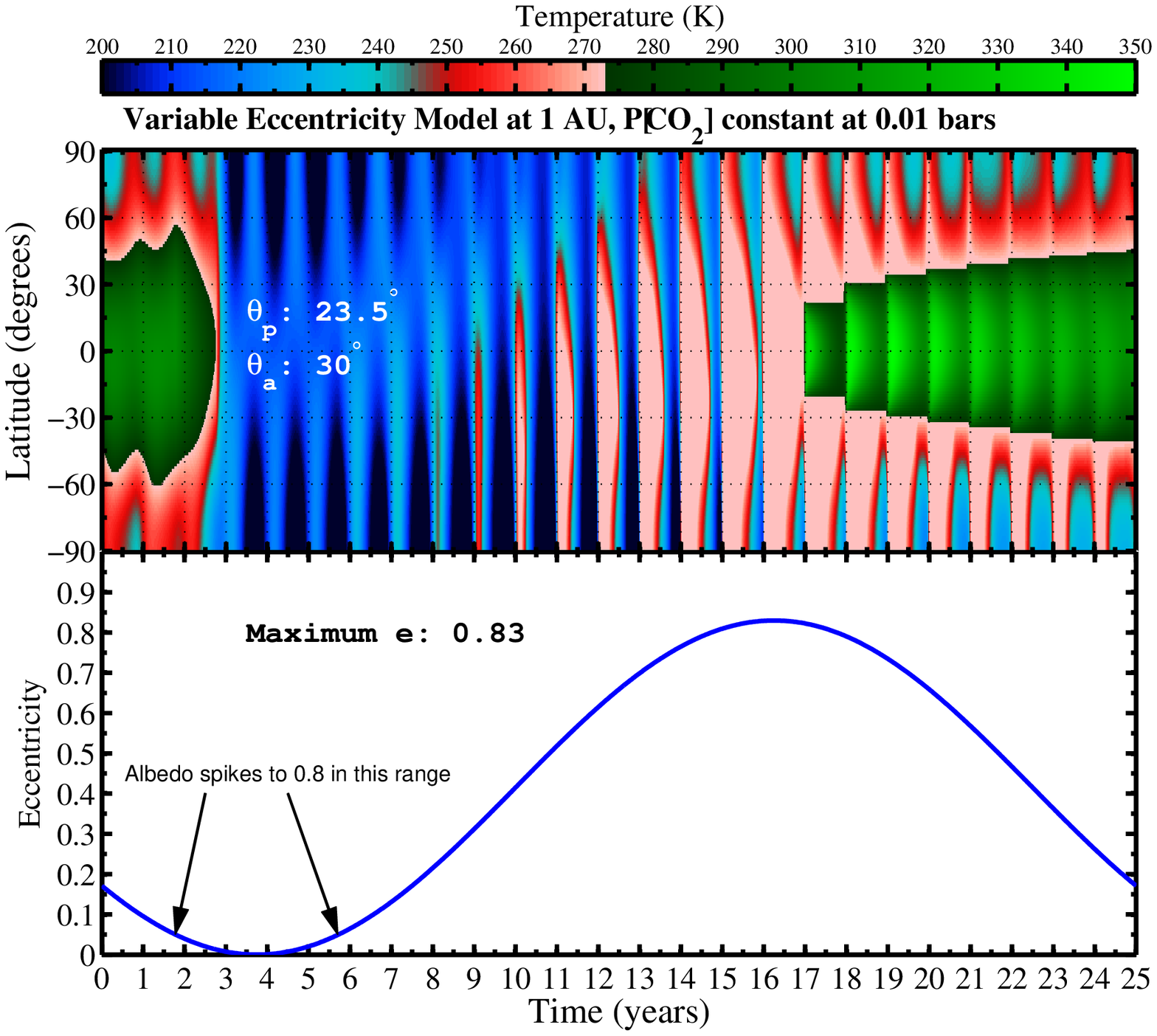}{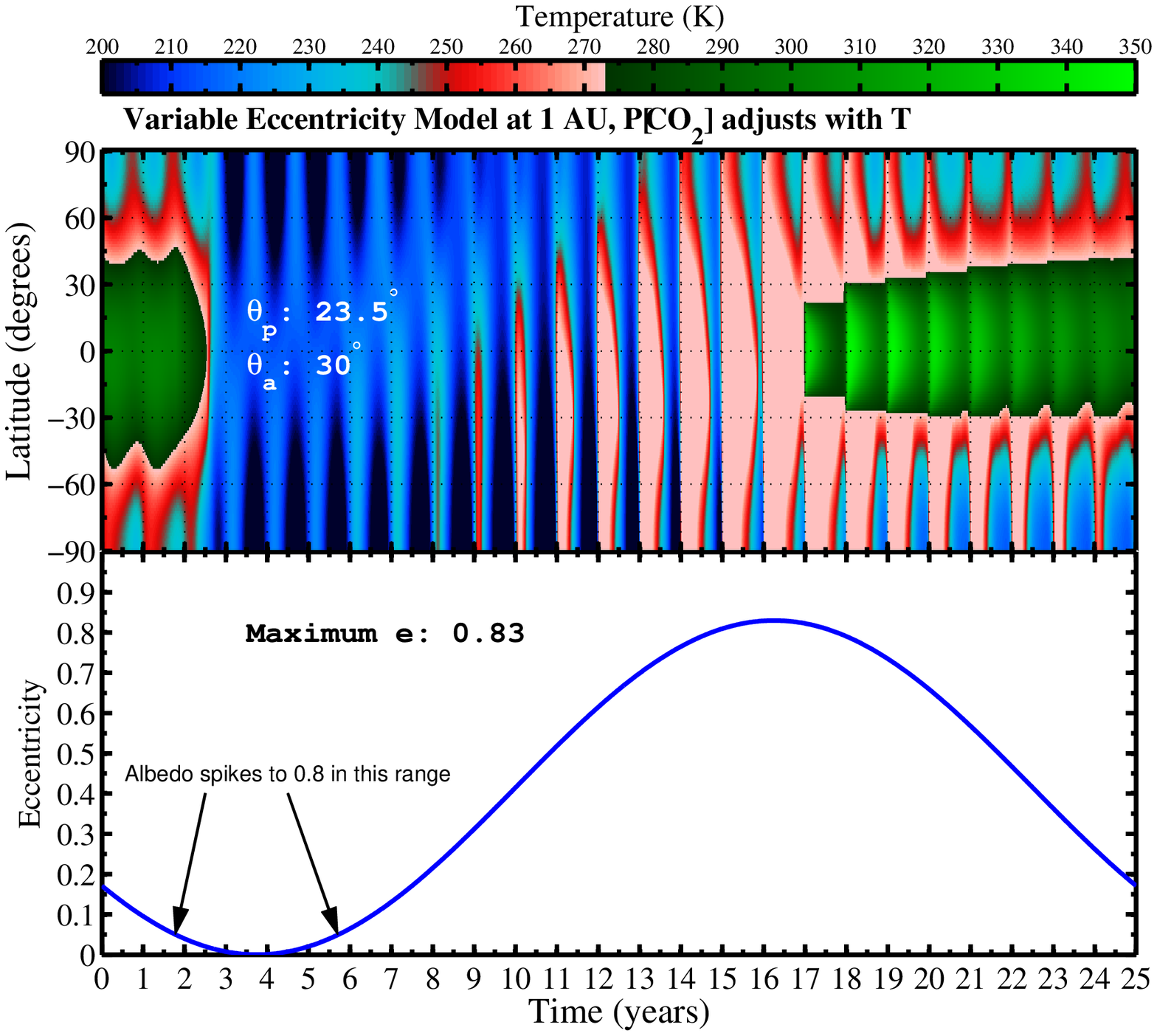}}
\centerline{\plottwo{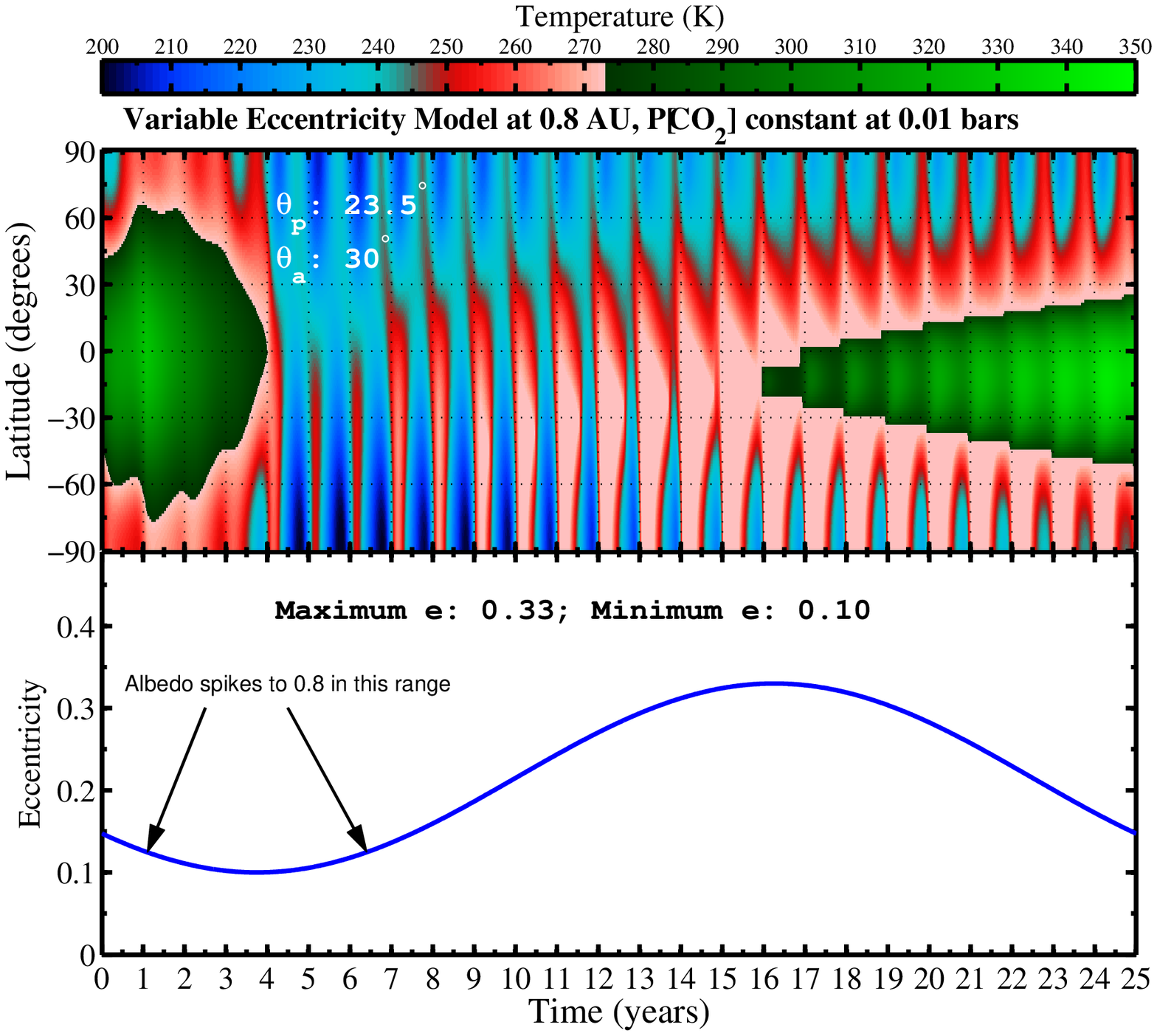}{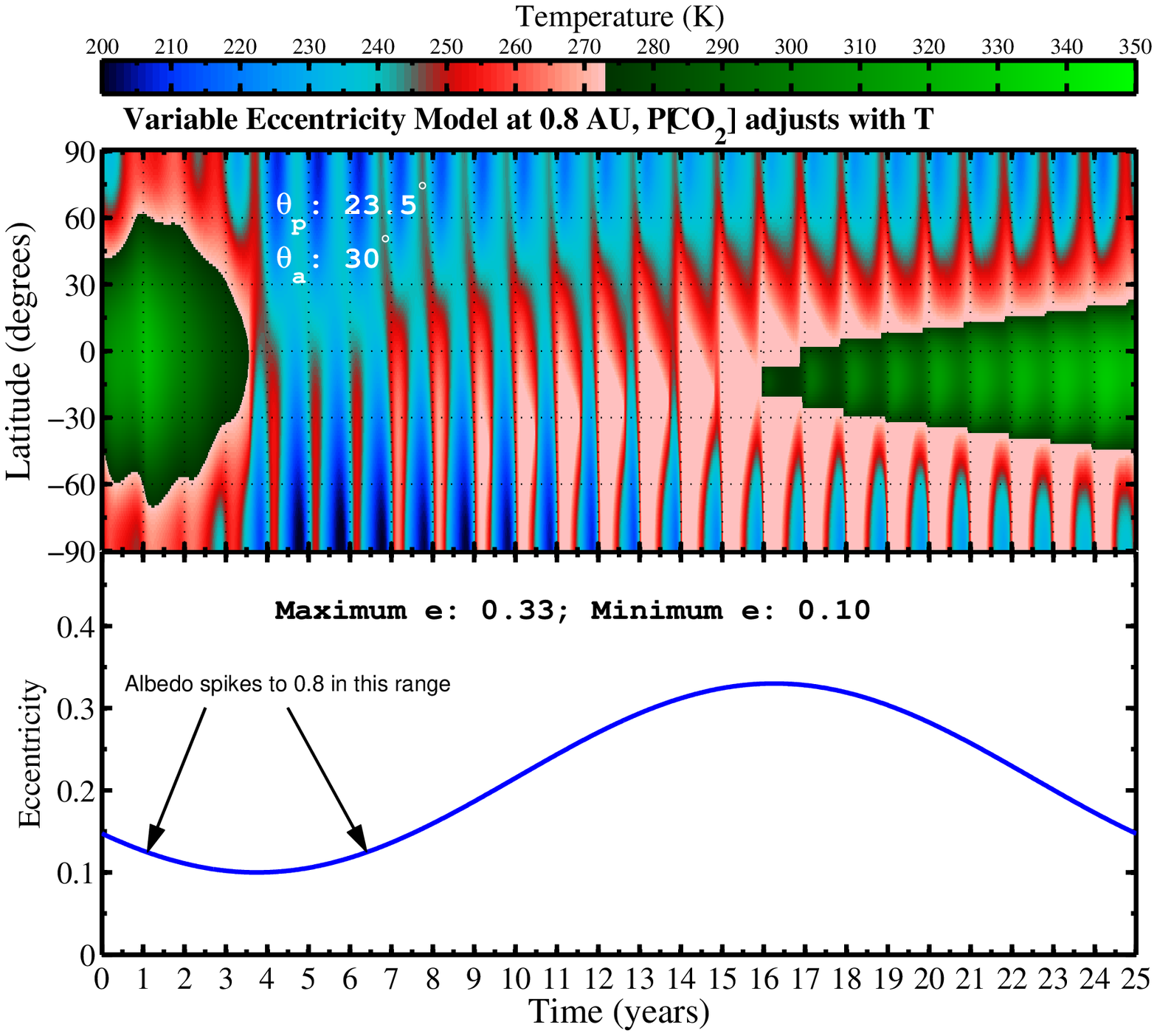}}
\caption{Compressed Milankovitch-like evolution of eccentricity and
  temperature at 1~AU and at 0.8~AU.  Planets are initialized with
  warm equator and cold poles, similar to present-day Earth.  In the
  top row (1~AU), the model planets are the same as in
  Fig.~\ref{fig:tempmap}, except the eccentricity varies sinusoidally
  between 0 and 0.83 with a 25-year period, to simulate a
  time-acceleration (by a factor of $\sim$10$^2$ to $\sim$10$^4$) of a
  Milankovitch-like cycle.  When the eccentricity falls below 0.05,
  the planet's albedo spikes to 0.8, simulating a catastrophic event
  that plunges the planet into a snowball state, with the latent heat
  prescription of \S\ref{ssec:coldstart}. In the bottom row (0.8~AU),
  the eccentricity varies between 0.1 and 0.33, also with a 25-year
  period.  {\bf Left:} CO$_2$ partial pressure is held fixed at
  0.01~bars.  As in the left panel of Fig.~\ref{fig:tempmap}, these
  planets do not establish a temperate equilibrium.  {\bf Right:}
  CO$_2$ partial pressure varies with temperature.  Here, temperature
  increases are muted by reduced greenhouse effect once the ice-cover
  has melted somewhere.}
\label{fig:tempmap_eccen}
\end{figure}

On a real planet with an extreme Milankovitch-like cycle of
eccentricity oscillations, the climate never quite reaches a stable
year-to-year equilibrium, because the eccentricity is constantly
changing.  Still, even for the most extreme cases explored in
\S\ref{sec:source}, the changes in eccentricity are very small from
one year to the next, since the periods of eccentricity oscillation
are $\gsim$10$^3$~years.  As a numerical experiment to probe what can
happen when eccentricity is varied as a climate model runs, we
compress two Milankovitch-like eccentricity cycles into 25 years.  One
cycle is for a planet at 1~AU, and the other is for a planet at
0.8~AU.  For simplicity, we set the eccentricities of both model
planets to vary sinusoidally.  The 1-AU planet has an extreme
Milankovitch-like cycle, with eccentricity varying from 0 to 0.83: $e
= 0.83 \{1 + 0.5\sin[(t-10)2\pi/25]\}$, and the 0.8-AU planet has a
more typical cycle, with eccentricity varying from 0.1 to 0.33.  We
start both planets with temperate conditions, warm equator and cold
poles ($T_{\rm init}[\lambda] = 240+60\cos^2\lambda$).  Furthermore,
when the eccentricity is below 0.05 for the 1-AU planet and below
0.125 for the 0.8-AU planet, we set the albedo to 0.8, simulating a
catastrophic event that plunges the planet into a snowball state in
which infrared cooling is governed by $I_{\rm WK97}$, initially with
constant $p_{\rm CO_2}=0.01$~bars.  Figure~\ref{fig:tempmap_eccen}
displays the results of four such model runs (two for planets at 1~AU,
two for planets at 0.8~AU; all planets have 23.5$\degr$ polar
obliquity).  In the left panels, the CO$_2$ concentration is held
constant at 0.01~bars throughout; in the right panels, the CO$_2$
concentration drops according to the prescription in
\S\ref{ssec:coldstart} once a region of surface has remained above
freezing for an entire year.  As in Fig.~\ref{fig:tempmap}, the
negative feedback of the thermostat's turning on mutes the positive
feedback of the ice's melting, resulting in a smaller fraction of the
ice-cover melting within the cycle.

\subsection{Discussion}
\label{ssec:disc}
In \S\ref{ssec:coldstart_results} and \S\ref{ssec:milank}, we
presented examples of planets with eccentricities between 0.1 and
0.86, and with eccentricity oscillation cycles of magnitude 0.83 and
0.23.  Although oscillations greater than 0.5 in a terrestrial
planet's eccentricity (such as those in Fig.~\ref{fig:tempmap_eccen}
and some of those in Table~\ref{ta:required_ecc}) might be considered
extreme, oscillations of $\sim$0.3 may be routine.  Furthermore, a
planet's variable eccentricity need not always return to zero (see the
lower right panel of Fig.~\ref{fig:e-t}), so our galaxy might host
many planets whose eccentricities vary between, say, 0.1 and 0.4.
Given this, it is worth keeping in mind that the snowball cases
considered above might be extreme, as well.  Some climate simulations
suggest that the Earth's snowball state(s) (if in fact it went through
any) might have left an equatorial band of open water, perhaps because
intense winds (driven by the sharp temperature contrast between frozen
and unfrozen tropical ocean) stirred up the typically $\sim$50~m
wind-mixed layer \citep{hartmann1994} to much deeper, tapping into the
large heat capacity of the deep ocean \citep{hyde_et_al2000}.  The
idea of a temperate region persisting throughout the (partial)
snowball state is also supported by the fact that life evidently
survived any such states, which might have been difficult if the whole
surface was covered with ice for $\gsim$10$^6$ years.  Much milder
(and more probable) eccentricity excursions could cause a planet in
either a partial snowball state or in an analog of the Earth's ice
ages to warm significantly.

In our models, we have held each planet's obliquity constant (at
23.5$\degr$ or 90$\degr$, depending on the model).  This assumption is
purely to limit the complexity of our modeling endeavor; it is not
motivated by astrophysical intuition or theory.  The architecture of a
planetary system affects not only a terrestrial planet's orbital
eccentricity, as considered in \S\ref{sec:source}, but also its
obliquity.  Since terrestrial exoplanets' obliquities might vary
significantly \citep{williams1998a, williams1998}, our assumption of
constant obliquity restricts the richness of the problem; the
evolution of climate and ice-distribution on a planet whose spin-axis
direction and orbital eccentricity are both varying wildly is much
more complex than we have considered.

Terrestrial exoplanets might someday be discovered that have
spectroscopic and photometric characteristics indicative of being
covered with snow and ice (and possibly of being shrouded in cold
CO$_2$ clouds) -- i.e., that seem to be in snowball states.  If such
planets are found, analyzing the full architectures of their planetary
systems will allow us to understand whether their eccentricities might
vary enough (and on what timescales) to help them to melt out of their
frozen states.  It might even be possible to determine whether a
planet is in the process of increasing (or decreasing) its orbital
eccentricity, and therefore whether it could be in the process of
melting (or freezing), which might, at some point, be testable through
direct observations.

\section{Origin of Earth-like Exoplanet Eccentricities}
\label{sec:source}
For the purposes of this study, we are interested in the eccentricity
evolution of Earth-like planets, in terms of both the amplitude of
oscillation and the period of oscillation.  Given the importance of
resonances and the potential for chaos, it is impossible to understand
the detailed dynamical evolution of a planetary system without full
knowledge of the system's architecture.  However, we can make simple
assumptions to get an idea of the range of possibilities.  We
therefore restrict ourselves to the simple case of an Earth analog at
1~AU evolving under the perturbations from a single massive body, in
most cases a Jupiter-mass giant planet. To study these simple,
two-planet systems we numerically integrated the orbits of several
systems to look at the range of outcomes and correlations between the
perturbing body's mass/orbit and the evolution of the Earth-like
planet.

We used two different numerical techniques to integrate our Earth
analog - single perturber systems.  First, to explore the range of
outcomes in coplanar systems, we integrate the orbit-averaged rates of
change for the planets' eccentricities $e$ and longitudes of
pericenter $\varpi$ using the equations of \citet{mardling+lin2002}
and \citet{mardling2007}.  Our second numerical technique was to
directly track the 3-dimensional orbits of the planets with the {\tt
  Mercury} symplectic integrator \citep{chambers1999}.  We used the
orbit-averaged method to study coplanar systems, and {\tt Mercury} to
verify the results of the orbit-averaged code and also to study
mutually inclined systems.  We evolved a variety of systems for 1 Myr,
in each case including an Earth-mass planet at 1 AU (initially on a
circular orbit), and neglecting the effects of general-relativistic
precession.  Our simulated systems were chosen not to make predictions
about specific known exoplanetary systems but rather to illustrate the
variety of eccentricity oscillations of terrestrial planets in
dynamically stable (albeit simplified) planetary systems, and to show
simple correlations between the properties of the perturbing body and
the eccentricity of the Earth analog.

\begin{figure}
\epsscale{0.6}
\centerline{\plottwo{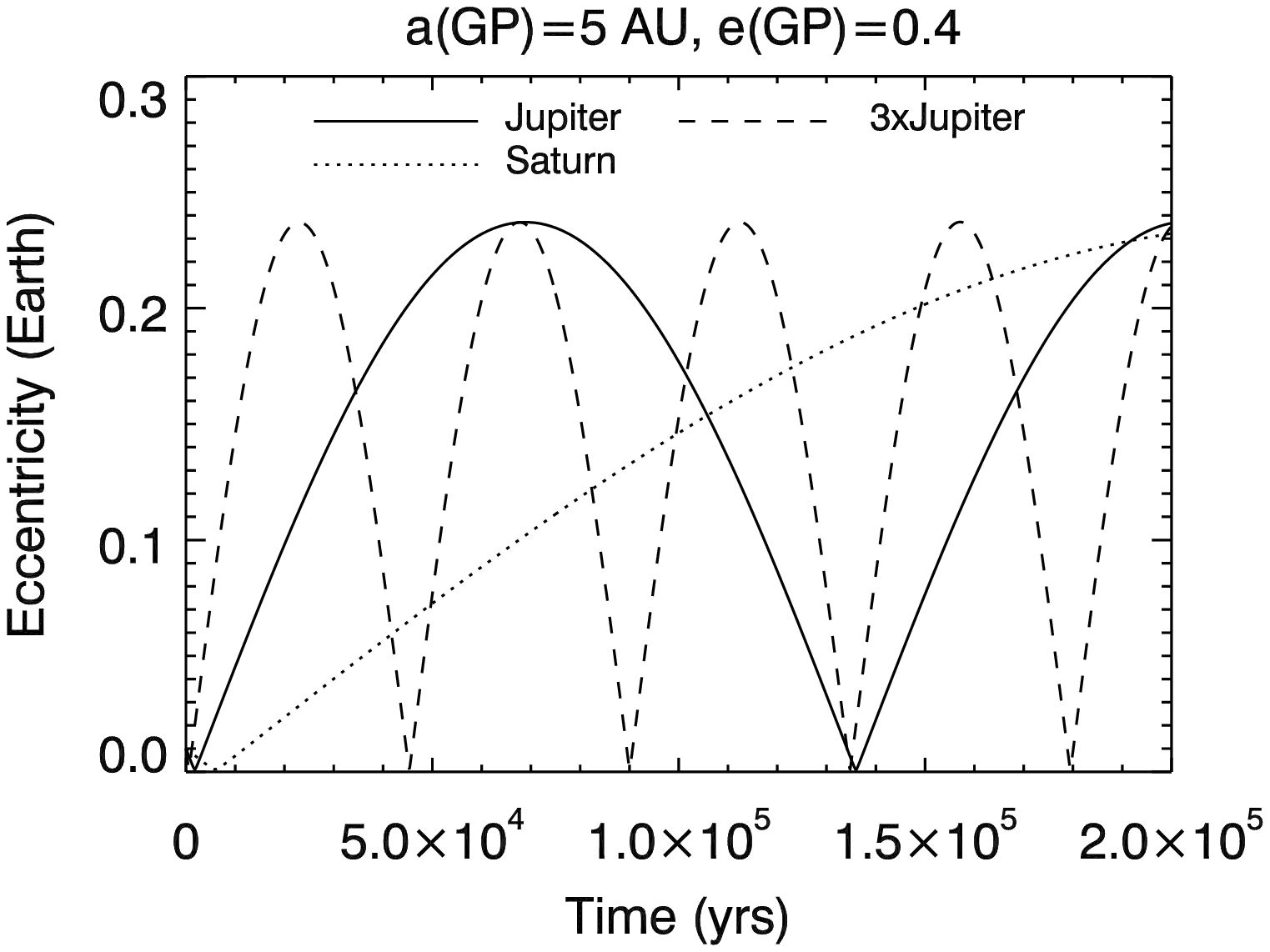}{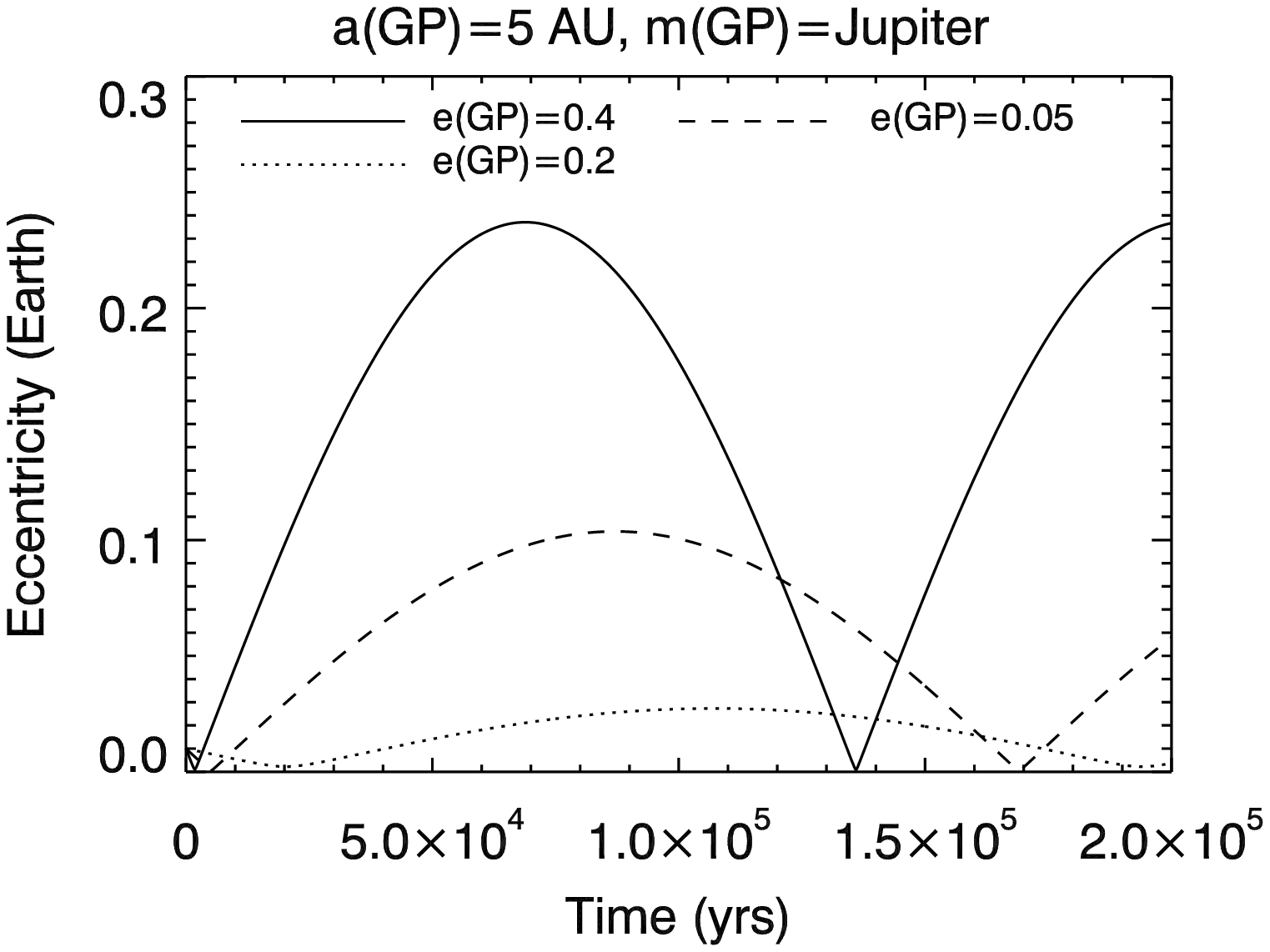}}
\centerline{\plottwo{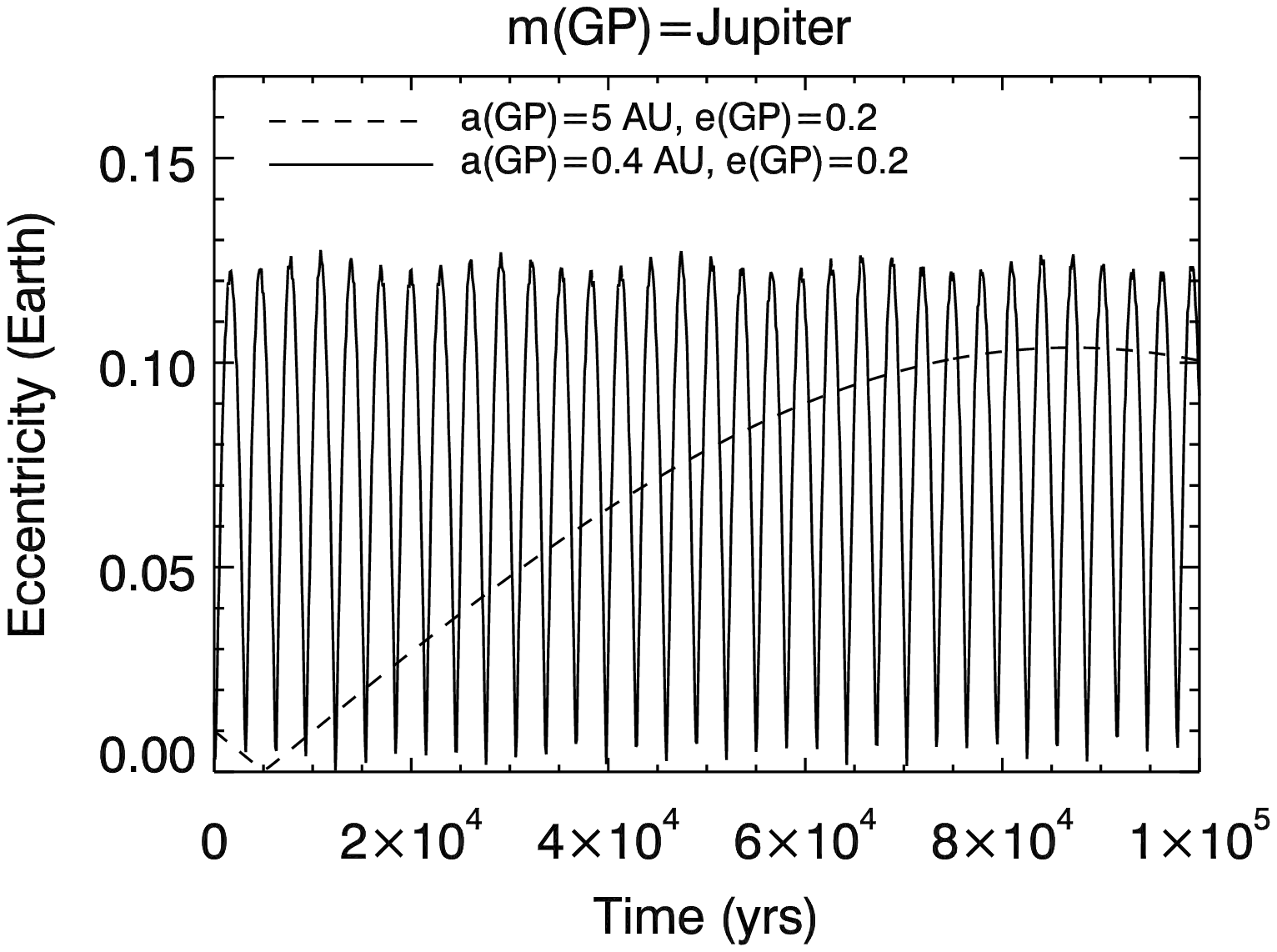}{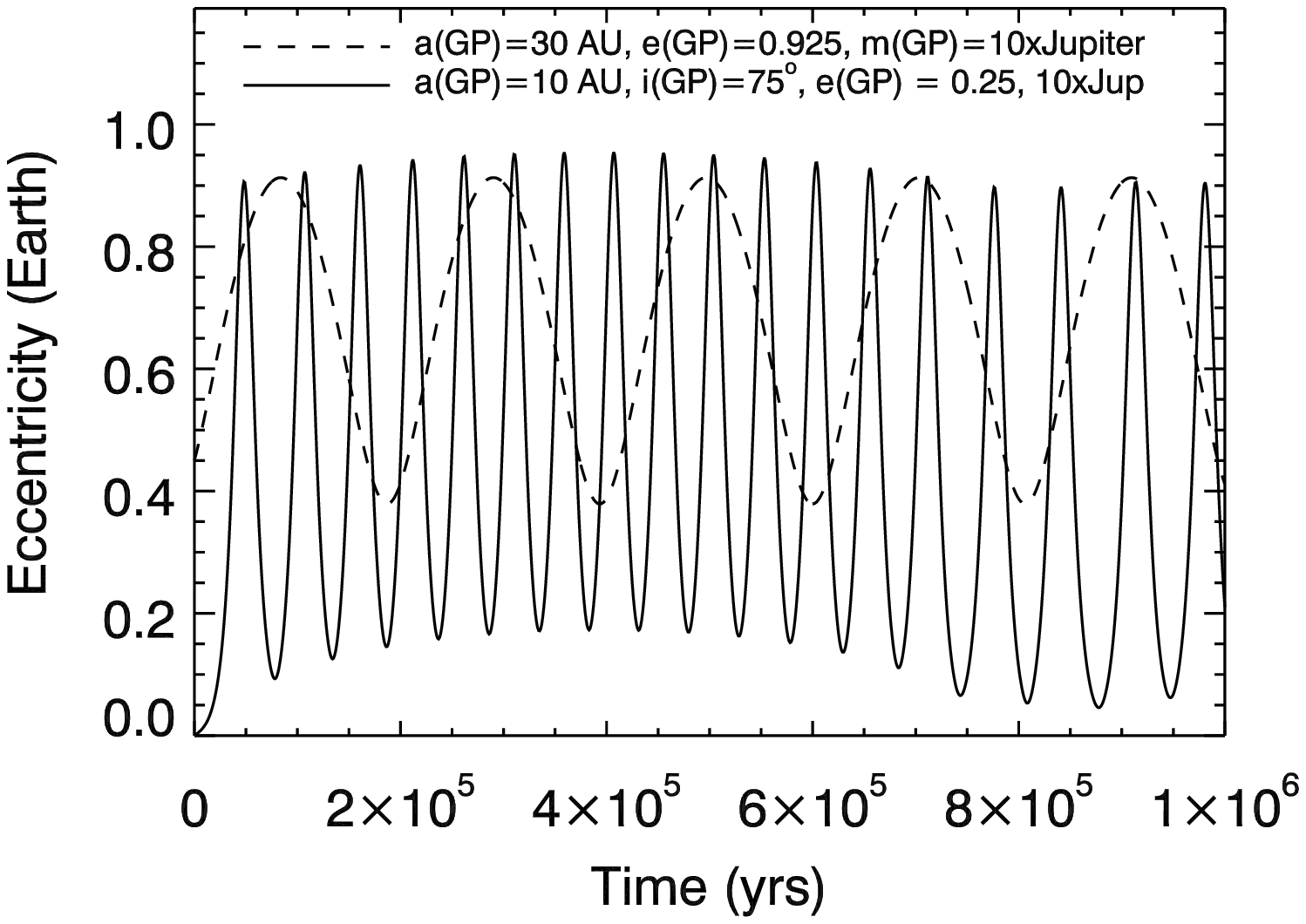}}
\caption{Eccentricity evolution of an Earth-mass planet at 1 AU under
  the influence of a range of giant planet masses and orbits, labeled
  by the giant planet ($GP$) semimajor axes $a$, eccentricities $e$,
  and masses $M$.  {\bf Top left:} Effect of changing the giant planet
  mass between Saturn's mass and 3$\times$ Jupiter's mass for the case
  of $a_{GP} = 5 {\rm AU}, e_{GP} = 0.4$.  {\bf Top right:} Effect of
  changing the giant planet eccentricity between 0.05 and 0.4 for the
  case of $a_{GP} = 5 {\rm AU}, M_{GP} = M_{Jup}$.  {\bf Bottom left:}
  Two cases with similar eccentricity amplitudes but very different
  planetary system architectures, although both with $M_{GP} =
  M_{Jup}$: $a_{GP} = 0.5 {\rm AU}, e_{GP} = 0.1$ (solid line) and
  $a_{GP} = 5 {\rm AU}, e_{GP} = 0.4$ (dashed line). {\bf Bottom
    right:} An extreme case, with $a_{GP} = 30 {\rm~AU}, e_{GP} =
  0.925$, and $M_{GP} = 10 \, M_{Jup}$ (dashed line), and with
  $a=10$~AU, $e_{GP} = 0.25$, $M_{GP} = 10\, M_{Jup}$, and
  $i_{GP}=75\degr$ (solid line). }
\label{fig:e-t}
\end{figure}

The results of our integrations are summarized in
Fig.~\ref{fig:e-t}.  The first three panels show the effect of a
particular system parameter: the giant planet mass (top left panel),
the giant planet eccentricity (top right panel) and the giant planet's
orbital distance (bottom right panel).  The amplitude of the
Earth-like planet's eccentricity oscillation is a function of the
giant planet's semimajor axis and eccentricity but is independent of
the giant planet's mass (top left panel).  However, the oscillation
frequency (often referred to as the secular frequency) increases
linearly with giant planet mass.  The amplitude of the Earth analog's
eccentricity oscillation decreases for lower giant planet
eccentricities, and the period of oscillation increases with lower
giant planet eccentricities (top right panel of Fig.~\ref{fig:e-t}).

The bottom left panel of Fig.~\ref{fig:e-t} shows two systems that
both induce eccentricity oscillations of about 0.1 in the Earth
analog, but on very different timescales of 3,000 and 170,000 years.
The two systems are markedly different in their architecture: the
fast-oscillating case contains a giant planet interior to the
Earth-like planet (giant planet at 0.5 AU with an eccentricity of
0.1); and the slow-oscillating case contains an exterior giant planet
(at 5 AU with an eccentricity of 0.4).

The first three examples from Fig.~\ref{fig:e-t} were chosen such that
the giant planets' orbits are consistent with the formation of an
Earth-like planet at 1 AU in systems with both inner
\citep{raymond_et_al2006, mandell_et_al2007} and outer
\citep{raymond2006} giant planets.  For the fourth example, we
searched for extreme cases that would be dynamically stable, yet
induce very large amplitude oscillations in the Earth analog's
eccentricity.  One potential source of large eccentricities comes from
a consideration of inclinations called the Kozai mechanism
\citep{kozai1962}.  For a star-planet-companion system, where the
companion can be stellar or planetary, an inclination between the
planet and companion's orbits $I$ larger than 39.2$^\circ$ can induce
eccentricity (and inclination) oscillations in the planet's orbit,
with a maximum eccentricity $e_{\rm max}$ of
\begin{equation}
e_{\rm max} \approx \sqrt{1 - \frac{5}{3} \cos^2(I)} \, ,
\label{eq:emax}
\end{equation}
and with a periodicity $P_{\rm Koz}$ of roughly
\begin{equation}
P_{\rm Koz} = P_1 \left(\frac{M_\star+M_1}{M_2}\right) \left(
\frac{a_2}{a_1}\right)^3 \, \left(1 - e_2^2\right)^{3/2} \, .
\label{eq:Pkoz}
\end{equation}
In equation~\ref{eq:Pkoz}, $M_\star$ is the stellar mass, $a$ is
semimajor axes, and subscripts 1 and 2 refer to the planet and the
companion, respectively \citep{innanen_et_al1997, holman_et_al1997,
  ford_et_al2000}.  For reasonable estimates of the mutual inclination
between planetary and companion orbits, the majority of binary stars
should induce large eccentricity oscillations in any planets that form
around the primary star \citep{takeda+rasio2005}.

Two extreme cases are shown in the bottom right panel of
Fig.~\ref{fig:e-t}, representing large-scale eccentricity forcing from
two dynamical sources: secular forcing in the coplanar system, as seen
in the previous examples, and the Kozai mechanism.  The secular
forcing example is a system with a 10 Jupiter-mass outer planet at 30
AU with an orbital eccentricity of 0.925.\footnote{Note that three
  known planets (HD~4113b, HD~80606b, and HD~20782b) have orbits with
  eccentricities of 0.9 or larger \citep{tamuz_et_al2008,
    Naef_et_al_2001, jones_et_al2006}.}  In this case, the Earth
analog started with an eccentricity of 0.45; thus, the ``backstory''
for this system involves an impulsive perturbation to the Earth analog
that increased its eccentricity considerably, followed by stable
secular interactions with the giant planet.  The Kozai mechanism
example contains a 10 Jupiter-mass planet orbiting at 10 AU with an
eccentric and inclined orbit: eccentricity of 0.25 and an inclination
of 75$^\circ$ with respect to the Earth-like planet's starting orbital
plane.  In both cases, the eccentricity of the planet at 1 AU reaches
0.9 or higher on a timescale of one to several hundred thousand years.
However, it is important to note that the inclination variations of
these planets are markedly different.  The system for which the Kozai
effect is important induces very large variations in the Earth
analog's inclination, going so far as to sometimes rotate in a
retrograde sense with respect to its initial orbital plane.  In
contrast, the secularly forced system remains virtually coplanar.

Resonances and the proximity of the giant planet are both of great
importance to the eccentricity evolution of an Earth-like planet.  To
test these effects we performed a small number of integrations of
systems containing one giant planet and a disk of massless test
particles whose behavior represents an ensemble of hypothetical
systems with Earth-like planets at different orbital distances.  We
used the hybrid version of the {\tt Mercury} integrator, taking care
to choose a step size small enough to resolve our innermost orbits at
least 20 times.  Figure~\ref{fig:emax} shows the maximum eccentricity
reached by test particles in the terrestrial planet zone for two
different orbital configurations of a Jupiter-mass planet, one case
with the giant planet exterior to 1~AU (shown in black) and one case
with the giant planet interior to 1~AU (shown in grey).  As expected,
the eccentricity decreases farther from the giant planet, in opposite
radial directions for the two cases, which induce similar
eccentricities in planets at 1 AU.  The ``blips'' in each curve
correspond to individual mean motion resonances with the giant planet,
which are stronger than in the Solar System because of the giant
planets' eccentricities (e.g., \citealt{murray+dermott2000}).  The 4:1
resonance with the inner giant planet is visible at 1 AU.

To summarize, the eccentricities of Earth-like planets are determined
either during their formation or the subsequent dynamical evolution of
the planetary system -- in many cases this will include a dynamical
instability among the giant planets.  Once the system has settled
down, the eccentricity of terrestrial planets is determined by the
influence of massive perturbers such as giant planets and binary
companions, via secular, resonant, and Kozai forcing.  For simple
systems with a single perturbing planet, the eccentricity oscillation
amplitude and period are determined by the giant planet's mass,
semimajor axis, and eccentricity.  For large mutual inclinations, the
Kozai mechanism is governed by the inclination between the planetary
and companion orbits, and the mass of the companion.  The main
difference between systems with giant planets interior to the
terrestrial planets and systems with giant planets exterior to the
terrestrial planets is that the eccentricity oscillation timescale
(the secular timescale) is much shorter for close-in giant planet
systems.  In systems that do not form any giant planets and have no
distant companions, there is no obvious mechanism by which large
eccentricities should be induced in Earth-like planets.  We therefore
expect planets in such systems to have relatively circular orbits
\citep{raymond_et_al2007}.

The so-called dynamical habitability of planetary systems has been
studied via the stability of test particles in the habitable zone of
the known extra-solar systems (e.g., \citealt{menou+tabachnik2003,
  barnes+raymond2004, rivera+haghighipour2007}).  This is an
interesting exercise because it can help observers by constraining the
orbital location of additional planets (see
\citealt{barnes_et_al2008}) and can also test the eccentricity
variations of stable test planets.  Such stable planets could indeed
exist in these systems below the detection threshold and represent
Earth analogs.

\section{Conclusion and Summary}
\label{sec:conc}
We have demonstrated that suitably structured exoplanetary systems
could contain Earth-like planets that undergo exaggerated versions of
the Earth's eccentricity Milankovitch cycles.  Plausible arrangements
of giant and terrestrial planets could lead to an Earth-like planet's
eccentricity oscillating between 0 and more than 0.9, on time-scales
of a few times $10^3$ years to a few times $10^5$ years.  Oscillations
such as these could have important consequences for climate and
habitability.

A companion paper (D10) investigates the influence of a temperate
terrestrial planet's eccentricity on its habitability.  Here, we have
examined how eccentricity could affect a globally frozen terrestrial
planet by introducing an algorithm to allow our EBM to treat the
melting of a global ice layer.  We found that increasing the
eccentricity of an initially-frozen model to sufficiently high values
can cause the model to melt out of the snowball state.  Such high
values of eccentricity, however, might place a planet at risk of
heating up to temperatures greater than 400~K (perhaps even undergoing
a runaway greenhouse effect) if there is not a concomitant negative
feedback, such as from a chemical thermostat similar to the Earth's
carbonate-siicate weathering cycle.  Because the climatic influence of
increased eccentricity depends on features of a planet's geochemistry
that might differ from planet to planet, our results should not be
taken as the final word on precisely what values of eccentricity would
be required.  Nevertheless, the principle remains that if a planet's
eccentricity is excited to large values, it is conceivable there are
situations in which this might actually increase, not decrease, its
habitability.

The longterm habitability of a planet depends on the properties of its
star, on the properties of its own orbit, spin, and geochemistry, and
also on the arrangement of other companion planets.  When, in the
coming years, we find Earth-sized planets at orbital separations that
could be consistent with habitable climates, it will be important to
study the architectures of those systems, so as to determine what
kinds of Milankovitch-like cycles might govern the longterm climatic
habitability of such worlds.

\acknowledgments

We thank Kristen Menou, Frits Paerels, Adam Burrows, Ed Turner, Scott
Tremaine, and Brian Jackson for helpful discussions.  We thank an
anonymous referee for comments and suggestions that materially
improved the manuscript.
We acknowledge the use of the Della supercomputer at the TIGRESS High
Performance Computing and Visualization Center at Princeton
University.  D.S.S. acknowledges support from NASA grant NNX07AG80G
and from JPL/Spitzer Agreements 1328092, 1348668, and 1312647.  S.N.R.
acknowledges funding from NASA Astrobiology Institutes' Virtual
Planetary Laboratory lead team, supported by NASA under Cooperative
Agreement No. NNH05ZDA001C.  This research was supported in part by
the National Science Foundation under Grant No. PHY05-51164.


\bibliography{biblio.bib}
\newpage

\clearpage

\begin{table}[ht]
\small
\begin{center}
\caption{Parameter Values Over Land, Ocean, Ice} \label{ta:land_ice_ocean}
\begin{tabular}{l|cccc}
\hline
\hline
\multirow{2}{*} {Parameter} & {Thick Ice}           & {Thin Ice}                & \multirow{2}{*} {Land} & \multirow{2}{*} {Ocean} \\
                            & {($T \le 263 \rm~K$)} & {($263 < T < 273 \rm~K$)} &                        &                         \\
\hline
\rule {-3pt} {10pt}
$C$\tablenotemark{a}              & $2 C_l$               & $9.2 C_l$                 & $C_l = 5.25\times 10^9 \rm~erg~cm^{-2}~K^{-1}$ & $40 C_l$ \\[0.2cm]
$\tau_{\rm diff}$\tablenotemark{b} & 7 months              & 2.8 years                 & 3.7 months         & 12 years \\[0.2cm]
$\tau_{\rm rad}$\tablenotemark{c}  & 3 months              & 1.1 years                 & 1.4 months         & 5 years \\[0.2cm]
$K_{yy}$\tablenotemark{d}        & $2.1\times 10^{10} \rm~cm^2~s^{-1}$ & $4.5\times 10^{9} \rm~cm^2~s^{-1}$ & $4.2\times 10^{10} \rm~cm^2~s^{-1}$ & $1.0\times 10^{9} \rm~cm^2~s^{-1}$ \\[0.2cm]
$\mathcal{K}^*$\tablenotemark{e}  & 1.6                   & 0.3                       & 3                  & 0.08 \\[0.2cm]
$\mathcal{R}^*$\tablenotemark{f}  & 4                     & 0.9                       & 8                  & 0.2 \\[0.2cm]
  \tableline
\end{tabular}
\tablenotetext{a}{$C$ is the heat capacity.}
\tablenotetext{b}{$\tau_{\rm diff} \equiv C/D$ is the diffusion time, where $D = 5.394 \times 10^2 \rm~erg~cm^{-2}~s^{-1}~K^{-1}$.}
\tablenotetext{c}{$\tau_{\rm rad} = (C/\sigma) {T_0}^{-3} = (1.4{\rm~months}) (C/C_l) (T/{T_0})^{-3}$ is the radiative time, taking $T_0$ to be an Earth-like 290~K.}
\tablenotetext{d}{$K_{yy} \equiv {R_\earth}^2/\tau_{\rm diff}$ is the eddy diffusion coefficient.}
\tablenotetext{e}{$\mathcal{K}^* \equiv P_{\rm orb}/\tau_{\rm diff}$ is the nondimensional diffusivity of equation~(\ref{eq:nondim}).}
\tablenotetext{f}{$\mathcal{R}^* \equiv P_{\rm orb}/\tau_{\rm rad}$ is the nondimensional thermal elasticity.}
\end{center}
\end{table}

\begin{table}[ht]
\small
\begin{center}
\caption{Eccentricity Required to Melt Cold-Start Models} \label{ta:required_ecc}
\begin{tabular}{ccccc}
\hline
\hline
Semimajor Axis & {$\theta_p$} & {$\theta_a$} & \multirow{2}{*} {IR Cooling Function} & \multirow{2}{*} {Required Eccentricity} \\
(AU)           &  (degrees)   & (degrees)    &                                       &                                         \\
\hline
\rule {-3pt} {10pt}
0.8            & 23.5         & 30           & $I_2$\tablenotemark{a}                & 0.39 \\
0.8            & 90           & 30           & $I_2$                                 & 0.51 \\
1.0            & 23.5         & 30           & $I_2$                                 & 0.86 \\
1.0            & 90           & 30           & $I_2$                                 & 0.79 \\
\hline
\rule {-3pt} {10pt}
0.8            & 23.5         & 30           & $I_{\rm WK97}[p_{\rm CO_2} = 0.01]$\tablenotemark{b} & 0.29 \\
0.8            & 90           & 30           & $I_{\rm WK97}[p_{\rm CO_2} = 0.01]$      & 0.44 \\
1.0            & 23.5         & 30           & $I_{\rm WK97}[p_{\rm CO_2} = 0.01]$      & 0.80 \\
1.0            & 90           & 30           & $I_{\rm WK97}[p_{\rm CO_2} = 0.01]$      & 0.74 \\
  \tableline
\end{tabular}
\tablenotetext{a}{$I_2$ is the cooling function from SMS08 and SMS09,
  and corresponds to present-day Earth-like greenhouse gas abundance
  (CO$_2$ partial pressure of $\sim$$3\times 10^{-4}$~bars).}
\tablenotetext{b}{$I_{\rm WK97}$ is the cooling function from
  \citet{williams+kasting1997}, and $p_{\rm CO_2}$ is the partial
  pressure of CO$_2$.}
\end{center}
\end{table}

\clearpage

\newpage
\clearpage

\newpage
\clearpage

\end{document}